\documentclass{aa}
\usepackage{graphicx}
\usepackage{supertabular}
\usepackage{natbib}

\bibpunct{(}{)}{;}{a}{}{,}

\begin{document}

\title{A spectroscopic event of $\eta$~Car viewed from different
directions: The data and first results\thanks{Based on observations
obtained with UVES at the ESO Very Large Telescope, Paranal,
Chile (proposals 70.D-0607(A), 71.D-0168(A), 072.D-0524(A))}}
\titlerunning{A spectroscopic event of $\eta$~Car}

\author{O. Stahl\inst{1} \and K. Weis\inst{2,}\thanks{Lise Meitner fellow}
  \and D.J. Bomans\inst{2}
  \and K. Davidson\inst{3}
  \and T.R. Gull\inst{4}
  \and R.M. Humphreys\inst{3}
  }

\offprints{O. Stahl, \email{O.Stahl@lsw.uni-heidelberg.de}}

\institute{Landessternwarte K\"onigstuhl, D-69117 Heidelberg, Germany
  \and Astronomisches Institut, Ruhr-Universit\"at Bochum,
  Universit\"atsstr. 150, D-44780 Bochum, Germany 
  \and
  Laboratory for Astronomy and Space Science,
  NASA-Goddard Space Flight Center, Code 681, Greenbelt, MD 20771, USA
  \and Astronomy
  Department, University of Minnesota, Minneapolis, MN 55455, USA 
}

\date{Received  / Accepted }

\abstract{We present spectroscopic observations with high spectral
resolution of $\eta$~Car as seen by the SE lobe of the Homunculus
nebula over the 2003.5 ``spectroscopic event''. The observed spectra
represent the stellar spectrum emitted near the pole of the star and
are much less contaminated with nebular emission lines than direct
observations of the central object.  The ``event'' is qualitatively
similar near the pole to what is observed in direct spectra of the
star (more equator-on at 45$^\circ$), but shows interesting
differences.  The observations show that the equivalent width changes
of H$\alpha$ emission and other lines are less pronounced at the pole
than in the line of sight. Also the absorption components appear less
variable. A pronounced high-velocity absorption is present near the
event in the He\,{\sc i} lines indicating a mass-ejection event. This
feature is also seen, but less pronounced, in the hydrogen
lines. He\,{\sc ii}$\lambda$4686 emission is observed for a brief
period of time near the event and appears, if corrected for light
travel time, to precede similar emission in the direct view. Our
observations indicate that the event is probably not only a change in
ionization and excitation structure or a simple eclipse-like
event.  }

\maketitle

\section{Introduction}
$\eta$~Car is one of the most massive and luminous stars of the
Galaxy. During its ``great eruption'' observed around 1843 it was one
of the brightest stars in the sky \citep[e.g.][]{1997ARA&A..35....1D}. Today
it is visually much fainter, largely because most of the UV and visual
light is absorbed by dust from the surrounding nebula and re-radiated in
the infrared \citep{1969ApJ...156L..45W,1995A&A...297..168C}. Its
bolometric luminosity is still very high
\citep{1999ecm..conf....6D}. The surrounding nebula is visible as a
bright reflection nebula, known as the Homunculus. Its
expansion has been detected already in the 1950s \citep[see][for
references]{1997ARA&A..35....1D}.  Recent proper motion measurements
indicate that most of this nebula was created during $\eta$~Car's
``great eruption''
\citep{1996AJ....112.1115C,1998AJ....116..823S,2001ApJ...548L.207M}.

The spectrum of $\eta$~Car is exceedingly complex. Seeing-limited
ground-based spectra of the star are strongly contaminated by the very
bright emission of the Weigelt blobs \citep{1986A&A...163L...5W} and
other ejecta close to $\eta$~Car. Only STIS spectroscopy can separate
the spectrum of the star from the nebular emission
\citep{1999ecm..conf..227D,1999ecm..conf..144G}.

The spectrum of $\eta$~Car is also known to be variable in
time. Marked changes in the high-excitation lines of $\eta$~Car have
been reported several times \citep[see references cited
by][]{1999ecm..conf..221D} before \citet{1996ApJ...460L..49D} first
noticed that these ``spectroscopic events'' recur in a 5.5-year
cycle. The exact nature of this cycle is not yet understood.  Most of
the very strong variations seen in the emission lines of ground-based
spectra of $\eta$~Car at the ``spectroscopic event''
\citep[e.g.][]{1984A&A...137...79Z} are in fact not due to spectral
variations of the central source but due to variations in the
condensations close to the center \citep{1999ecm..conf..116H}. A
spectroscopic scan of the Homunculus was done by
\citet{1992A&A...262..153H}. They could show that most the emission
lines in the outer Homunculus are scattered from the central region
and only some are intrinsic to the Homunculus.

The cycle is also observed as eclipse-like event in the X-rays
\citep{1999ApJ...524..983I}. In the visual range
\citep{2003IBVS.5477....1F} and the near-IR
\citep{2004MNRAS.352..447W} a short dip in the light-curve is also
observed.  In the radio range, structural changes and strong flux
changes of the nebula with the same period are observed
\citep{2003MNRAS.338..425D}.  \citet{2004AJ....127.2352M} observed the
cycle photometrically with HST and therefore could separate variations
of the star from variations of its environment. At present, the cause
of the cyclic variations is not known. While the exact periodicity
suggests a binary explanation, it is not obvious that such a scenario
can explain all the observed variations
\citep{1999ecm..conf..304D}. The spectroscopic changes are clearly not
due to a simple eclipse. Ground-based spectroscopic studies over a
complete cycle have been done
\citep{1998A&AS..133..299D,1999ecm..conf..243W,2000ApJ...528L.101D},
but give little information on the changes in the central source.
While the HST/STIS observations by \citet{2003ApJ...586..432S} and
more recent observations by the $\eta$~Car HST treasury team
\citep{Davidsonproc2004,Humphreysproc2004} provide the necessary
spatial resolution, the temporal sampling of these observations is not
as dense as desirable.

Due to the geometry of the nebula, the reflection in the Homunculus
allows us to view the spectrum of $\eta$~Car from different latitudes
above the star's surface.  This is of particular interest, since
HST/STIS long slit spectroscopy in 1998--2002 by
\citet{2003ApJ...586..432S} did show that the star is not spherical
and its wind is bimodal.  These results provided new insights with
crucial implications for the event, the nature of $\eta$~Car, its mass
loss history and bipolar ejecta.  \citet{2003ApJ...586..432S} find
that the wind from the polar regions is fast and dense while it is
slower and less dense at lower latitudes. The direct view on the
star corresponds to a stellar latitude of about 45$^\circ$. The strong
anisotropy of the stellar wind of $\eta$~Car is also indicated by
direct interferometric observations \citep{2003A&A...410L..37V} in the
near infrared, where the stellar wind can be directly
resolved. \citet{2003A&A...410L..37V} find that the stellar wind is
elongated in the direction of the axis of the Homunculus, indicating
an alignment of the axis of the Homunculus with the rotation axis of
the star.

Here we report ground-based VLT/UVES observations of the star's
reflected spectrum observed in the south-east lobe of the
Homunculus nebula. While these observations cannot match the spatial
resolution of the HST, they have a denser sampling, significantly
higher spectral resolution and excellent S/N-ratio. A short
description of the data set has been given by \citet{Weisproc2004}.

The Homunculus appears to be a very clumpy, thin bipolar shell with
hollow interior. \citet{2002MNRAS.337.1252S} notes that the near
infrared molecular hydrogen emission is consistent with the shell
thickness being about ten percent of its distance, $D$, from
$\eta$~Car.  Spectra recorded at the center of the foreground SE lobe,
known as ``FOS4'', appear to represent the stellar spectrum emitted
near the pole of the star
\citep{1995AJ....109.1784D,1999A&A...344..211Z}. They are much less
contaminated by nebular emission lines than seeing-limited ground-based
observations of the star directly. As has been shown by
\citet{2001ecom.conf...29R}, ground-based spectra of FOS4 can
therefore be used to study the spectral changes of the star itself
through the cycle.

The UVES project presented here is coordinated with STIS
observations. Some first results of the STIS \citet{Davidsonetal2005}
and UVES \citet{Weisetal2005}observations
have been reported already. The purpose of this paper is to present the data
and to give an overview of the spectroscopic variations. Due to the
large size and complexity of the data set and the strong spatial and
temporal variations, more detailed aspects will be discussed in
follow-up papers.

\section{The observations}

The data presented here are based on observations carried out in
service observing mode between December 2002 and March 2004 with
UVES, the {\bf U}ltraviolet and {\bf V}isual {\bf E}chelle {\bf
S}pectrograph at the Nasmyth platform B of ESO's VLT UT2 (Kueyen) on
Cerro Paranal, Chile. For all observations the standard settings DIC1,
346+580 and DIC2, 437+860 were used. The observed wavelength range
extends from about 3\,200 \AA\ to 10\,200 \AA\ except for small gaps due to
the space between the two CCDs of the detector mosaic at the red
channel. Since the main aim was the observation of the reflected light
of the Homunculus, the slit was offset from the star 2\farcs6 south
and 2\farcs8 east of the star and the slit was aligned along the SE
lobe of the Homunculus at a position angle of 160$^\circ$.  

The exact location of the slit is shown in the paper by \citet{Weisetal2005}.

The temporal sampling of our observations was quite dense close to the
event (about one spectrum every two weeks) and more sporadic further
away from the event. Unfortunately, $\eta$~Car is very low in the sky
between August and November which made it unobservable with
UVES\@. Therefore, the immediate recovery from the 2003.5 event was
not covered by our observations. STIS observations were accomplished
during this gap and will be discussed elsewhere.

In order to observe both the extremely strong H$\alpha$ emission and
fainter lines with good signal-to-noise ratio, two exposures with
different exposure times were obtained for the observations with the
DIC1, 346+580 setting. Typical exposure times are 10 seconds for the
short and 15 minutes for the long exposure. The slit width for the
observations was 0\farcs3 and 0\farcs4 in the red and blue range,
respectively, resulting in a spectral resolution of about 80\,0000.
The pixel scales and slit lengths were, respectively, 0\farcs246/pixel
and 7\farcs6 in the blue and 0\farcs182/pixel and 11\farcs8 in the red
arm.  The observational data were reduced and two-dimensional spectra
were extracted using mostly standard ESO pipeline software for UVES\@.
An exception was the order-merging procedure where the pipeline
software did not produce satisfactory results, mainly because the very
noisy edges of the echelle orders deteriorated the S/N in the
overlapping regions. Therefore, the order merging was carried out
using software developed at the LSW Heidelberg
\citep{1999oisc.conf..331S}. All spectral frames were converted to the
same (heliocentric) wavelength scale.

As a reference for the spectrum before the ``spectroscopic event'', we
also reduced and analyzed UVES spectra obtained during the UVES
commissioning (Dec. 21, 1999), which are released to the community and
have also been used by \citet{2001ecom.conf...29R}. They have been
obtained with the same wavelength setting as our spectra, but with a
different slit orientation. The position angle of the slit was
45$^\circ$, i.e.\ about perpendicular to the position angle of the
Homunculus and the slit was offset by 4\arcsec\ SE from the star. The
exposure time of the spectrum used was 60 sec. A figure with the exact
location of the slit is shown by \citet{Weisetal2005}.

Basic data for the spectra used are summarized in
\citet{Weisetal2005}.  The phase interval from 0.9
to 0.127 around the ``spectroscopic event'' is covered by our
observations. Unfortunately, due to observing constraints, there is a
significant gap between phase 0.016 and 0.074. Another smaller gap is
present between phase 0.946 and 0.970, but this is at a less critical
phase.

In order to study the temporal variations at specific positions in the
nebula, we have extracted one-dimensional spectra from the long-slit
spectra at five positions in the Homunculus. The part of the slit
which is close to the center shows a very complex spectrum and
significant amount of scattered light (due to seeing) from the central
source. For the present study, we therefore used only positions
located at center of the slit (about 2\hbox{$.\!\!^{\prime\prime}$}6
south and 2\hbox{$.\!\!^{\prime\prime}$}8 east of the central star)
and in steps of 0\farcs95 outwards along the slit.  We call these
position ``1'' to ``5'' in the following.  Position ``2'' is the
brightest position and close to the center of the Homunculus, where
already \citet{1999ecm..conf..107H} noticed that a very clear stellar
spectrum can be observed. This position is also known as FOS4 in the
literature (The FOS4 position was originally defined as a 0\farcs5
region 4\farcs03 from the center at a position angle of
135$^\circ$. Correcting for the expansion of the Homunculus, it would
now be at distance of 4\farcs21).  Five or three rows of the CCD have
been averaged in the red and blue spectral range, respectively, in
order to increase the S/N-ratio of the extracted spectra. As reference
for the spectrum close to the object, we also extracted one position
close to the star, denoted as ``0''. The extracted CCD rows lines have
been corrected for the velocity shift (see below and
Tab.~\ref{tab_velo}) before averaging, with the exception of the
spectra at position ``0'', which have not been corrected.

For better visualization, we also present the data in the form of time
series spectra, where the spectra, using the routines described by
\citet{1996A&A...305..887K}, are interpolated along the time axis
to form a ``dynamical spectrum''  When we extract information from the
spectra, we do not use these interpolated spectra, but the extracted
spectra directly.

\section{Basic data}

\subsection{Period and phase}

Slightly different values of the zero point and period of the cyclic
variations are given in the literature. The differences are small and
not relevant for the present paper.  In the following we shall use the
period and zero point of \citet{2004ApJ...612L.133S}, defined by the
disappearance of He\,{\sc i}$\lambda$6678, of $P=$ 2022.1 days and
$T_{\rm 0}$ = 2\,452\,819.8.

\subsection{Velocity shifts}

\begin{figure}[ht]
\resizebox{\hsize}{!}{\includegraphics[angle=0]{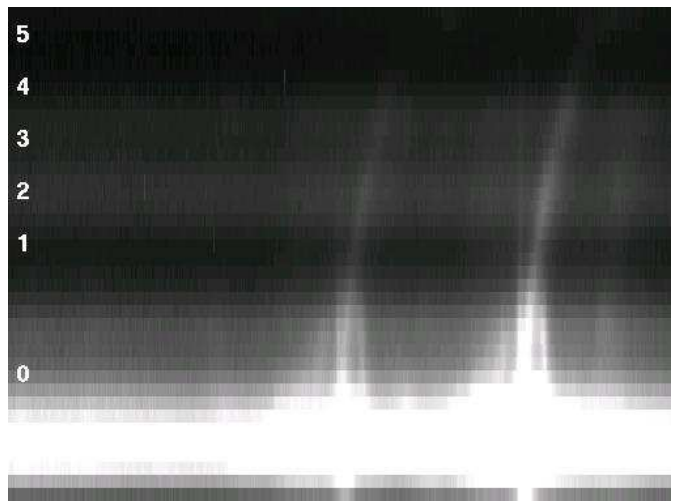}}
\caption{Long-slit spectrum around the [Fe\,{\sc
ii}]$\lambda\lambda$4277, 4287 lines.  The image covers 37 \AA\ in
wavelength and 9\farcs3 along the slit. Note the complex velocity
structure below the center of the slit.  The approximate extraction positions are
indicated. Here we discuss mainly the region above the center of the
slit.}
\label{fig_feii_longslit}
\end{figure}

The reflection in the Homunculus causes a wavelength shift and a
light-travel time delay.  The spectral lines in the reflected spectra
appear shifted in wavelength because of the expansion of the
Homunculus \citep{1961Obs....81...99T}. These wavelength shifts have
to be corrected in order to compare spectra extracted along the slit.
We measured the velocity shifts in spectra taken well before the
spectroscopic event in Dec.~2002, because they contain sharp forbidden
emission lines, which appear most suitable for this measurement. We
used several [Fe\,{\sc ii}] lines, which are relatively strong outside the
event. As an example we show in Fig.~\ref{fig_feii_longslit} a small
spectral range around the [Fe\,{\sc ii}]$\lambda$4287 line.

The heliocentric radial velocity has been measured for several lines
and averaged. The mean values have been fitted by a polynomial as a
function of the distance from the slit center. The polynomial values
for several positions along the slit are given in
Table~\ref{tab_velo}. The errors are of the order of 5 km
sec$^{-1}$. For the velocity correction we used the polynomial values
at the individual CCD rows to correct each row individually.  This
avoids the smearing due to different velocity shifts when averaging
CCD rows.

In order to determine the velocity shift due to reflection, these
velocity have still to be corrected for the velocity of the emitting
gas in the direction from the star to the scattering surface.  The
systemic velocity of $\eta$~Car has been estimated to $-7\pm10$ km
sec$^{-1}$ by \citet{1997AJ....113..335D} by measuring the velocities
of O stars near $\eta$~Car. This is in agreement with more recent
measurement of \citet{2004MNRAS.351L..15S}, who used H$_2$
measurements of the Homunculus nebula to obtain $-8.1\pm1$ km
sec$^{-1}$. The [Fe\,{\sc ii}] lines likely originate in compact
emission regions close to the central star, the so-called ``Weigelt
blobs''. These regions have a complex velocity structure
\citep{2004MNRAS.351L..15S} and are moving in the equatorial plane of
the nebula \citep{1997AJ....113..335D}. Their velocities towards the
SE lobe, which are relevant here, are probably small. We therefore
used the heliocentric velocity shifts without further correction. All
velocities quoted below and all extracted one-dimensional spectra are
corrected for this velocity shift.

\begin{table}
\caption{Heliocentric radial velocities of [Fe\,{\sc ii}] at various positions
lines along the slit. The offset is measured relative to the center of the
slit. Offsets relative to the center of the $\eta$~Car are also given. 
The errors are of the order of 5 km sec$^{-1}$. The velocities
at the centers of the extraction positions are also given.}
\label{tab_velo}
\begin{tabular}{lllll}
offset [$^{\prime\prime}$] & vel. [km sec$^{-1}$] 
& RA [$^{\prime\prime}$] & dec. [$^{\prime\prime}$]& pos. index\\
\hline
1 &    80  & 2.95 & 3.75   \\
1.5 &  92  & 3.10 & 4.20 \\
2 &   112  & 3.30 & 4.70   \\
2.5 & 141  & 3.45 & 5.15 \\
3 &   178  & 3.65 & 5.60  \\
3.5 & 224  & 3.80 & 6.10 \\
4 &   279  & 4.00 & 6.60 \\
4.5 & 341  & 4.15 & 7.00 \\
\hline
0    &  80 & 2.60 & 2.80 & ``1'' \\ 
0.95 &  80 & 2.95 & 3.70 & ``2'' \\
1.90 & 107 & 3.25 & 4.60 & ``3'' \\
2.85 & 166 & 3.60 & 5.50 & ``4'' \\
3.80 & 255 & 3.90 & 6.40 & ``5'' \\
\end{tabular}
\end{table}  

\subsection{Time delay}

The light-travel time delay is caused by the difference in paths
between direct line of sight and the longer path from the emitter to
the scattering surface to the observer. This time delay can be related
to the velocity shift. In the simplest model of the Homunculus nebula,
the time delay depends linearly on the velocity shift.  This model
assumes that the expansion velocity of the Homunculus is unchanged
since the eruption and the Homunculus is a hollow reflection nebula.

\citet{1987A&A...181..333M} derived the following expression for the
velocity shift caused by the reflection at the surface of the nebula:

\begin{equation}
v_\mathrm{shift} = (1 - \cos \theta)  v_\mathrm{dust}
\end{equation}

The time delay, on the other hand is given by:

\begin{equation}
t_\mathrm{delay} = (1 - \cos \theta) D_\mathrm{Homunculus} / c 
\end{equation}

Assuming a constant expansion velocity of the nebula, we have:

\begin{equation}
D_\mathrm{Homunculus} = v_\mathrm{dust} \times t_\mathrm{Homunculus}
\end{equation}

and the following relation between the velocity shift
$v_\mathrm{shift}$, the speed of light $c$ and the delay time
$t_\mathrm{delay}$ and the age of the Homunculus
$t_\mathrm{Homunculus}$ can be derived:

\begin{equation}
v_\mathrm{shift}/c = t_\mathrm{delay}/t_\mathrm{Homunculus}
\end{equation}

This relation directly links the velocity shift and the time delays.
For $t_\mathrm{delay}$ in days and $v_\mathrm{shift}$ in km
sec$^{-1}$, we get then for an expansion age of the Homunculus of 160 yr:

\begin{equation}
t_\mathrm{delay} = 0.195 \times v_\mathrm{shift} 
\end{equation}

I.e., for a velocity shift of 100 km sec$^{-1}$ in the direction of
FOS4, the light-travel delay time is about 20 days. 

The exact time and temporal evolution of the ``spectroscopic event''
may depend on stellar latitude. Before we can draw any conclusions
about this dependence, the observation times have to be corrected for
this light-travel time delay.

\section{Variations in different lines}

\subsection{Hydrogen Balmer and Paschen lines}

The detailed behaviour of the Balmer lines as observed at one
particular position in the Homunculus (FOS4) has already been
discussed by \citet{Weisetal2005}. Here we add some additional
information.  

\begin{figure}
\resizebox{\hsize}{!}{\includegraphics[angle=-90]{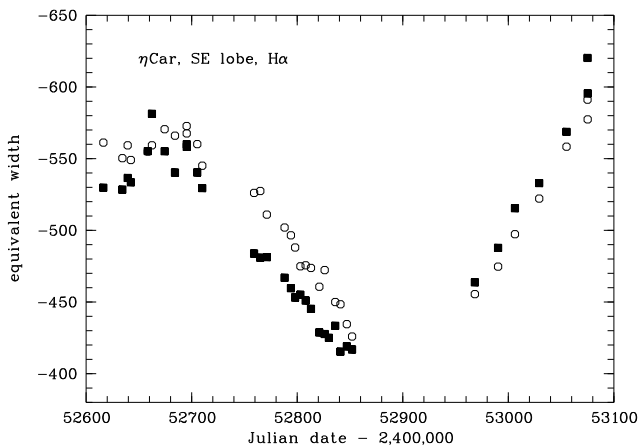}}
\caption{Equivalent width versus Julian Date for position ``2''
(filled symbols) and position ``5'' (open symbols). The curve at
position ``5'' seems delayed, but the time delay is smaller after
the event.}
\label{fig_eqw}
\end{figure}

Fig.~\ref{fig_eqw} shows the equivalent width of H$\alpha$, the
brightest observable feature, between 2002 and 2004. The equivalent
width has been measured between 6500 and 6620 \AA\@. The decrease of the
equivalent width during the event is clearly visible. The decrease of
equivalent width during the event appears later at position ``5'', as
expected because of the longer time delay.  The increase after the
event is less delayed, however, indicating that the delay is not only
a light travel effect. In spite of the significant difference in line
profiles, the equivalent width does not change strongly along the
slit. The equivalent width is, however, always significantly smaller
(by almost a factor of two) than observed directly on the star
\citep{Davidsonetal2005}. Also the amplitude of the variations is
smaller than in STIS spectra taken on the central source
\citep{2004AJ....127.2352M}. The decrease of the equivalent width
stars around JD 2\,452\,700 (phase 0.94) and the recovery is still
ongoing at the end of our observations at phase 0.127.

Seen directly of $\eta$~Car, the Balmer line absorption strongly
increases at the event \citep{Davidsonetal2005}. By contrast, the
Balmer line absorption in the polar spectra, as seen at FOS4 position,
always has P~Cygni absorption. 

\begin{figure}[ht]
\resizebox{\hsize}{!}{\includegraphics[angle=0]{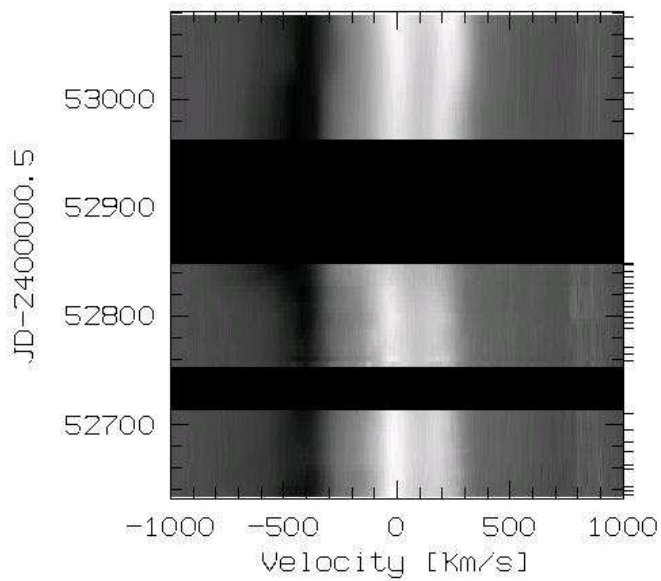}}
\caption{Time series spectrum of Paschen~8, as derived from the
observed spectra. Dates are given in JD-2\,400\,000.  The dates where
spectra have been taken are marked at the right. Note the strong
increase in edge velocity near JD 2\,452\,830. The red
bump is always very pronounced in the Paschen lines.}
\label{fig_P8_2D}
\end{figure}

\begin{figure}[ht]
\resizebox{\hsize}{!}{\includegraphics[angle=270]{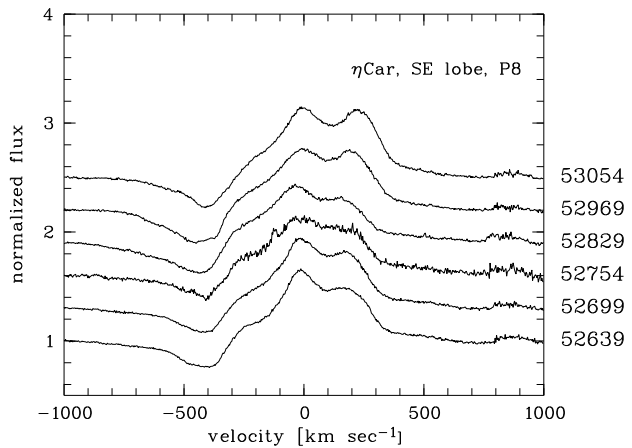}}
\caption{Selected spectra from the data set used in
Fig.~\ref{fig_P8_2D}. The spectra used have been taken close to the
beginning and end of the observing run and close to the gaps in the
observing runs. Note the strong increase in edge velocity close to the
``event'' and the strong red bump.}
\label{fig_P8_1D}
\end{figure}

The line profiles of the Paschen lines, as an exemplified by Paschen~8
in Figs.~\ref{fig_P8_2D} and~\ref{fig_P8_1D} behave in a similar
manner to the Balmer lines, but with a more pronounced red bump. The
FOS Paschen~8 line profiles (Figs.~\ref{fig_P8_2D},~\ref{fig_P8_1D})
have an absorption strength that increases rapidly at the event with
the edge velocity increasing significantly up to about $-$800 km
sec$^{-1}$.  The high velocities appear at JD 2\,452\,800 (phase
0.990) and are strongest around JD 2\,452\,830 (phase 1.005). The
increase of edge velocity is especially clear when comparing with
spectra obtained after the event. Before the event, the absorption
edge is very smooth and therefore ill-defined.  At the phase of
maximum edge velocity, the emission strength also seems to reach a
minimum value, although this is not completely clear due to the gap
shortly after after the event. The main peak has a velocity close to
zero. After the event, a hump is visible in the red at a velocity of
about $+200$ km sec$^{-1}$.The velocity of the bump is not constant
but varies between about $+150$ and $+220$ km sec$^{-1}$. The edge
velocity is also variable and reaches a maximum of at least $-800$ km
sec$^{-1}$.

\subsection{Fe\,{\sc ii}  lines}

\begin{figure}[ht]
\resizebox{\hsize}{!}{\includegraphics[angle=0]{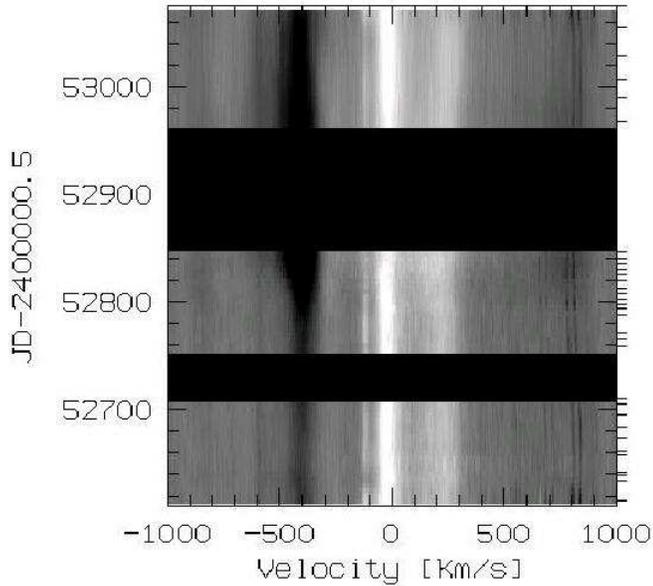}}
\caption{Same as Fig.~\ref{fig_P8_2D}, but for Fe\,{\sc
ii}$\lambda$6456.  The emission is relatively constant. The absorption
increases strongly at the event and remains strong for an extended
period. Recovery is not yet completed at the end of our observing
run. High-velocity absorption similar to the hydrogen lines is not
seen.}
\label{fig_feii_2D}
\end{figure}

\begin{figure}[ht]
\resizebox{\hsize}{!}{\includegraphics[angle=270]{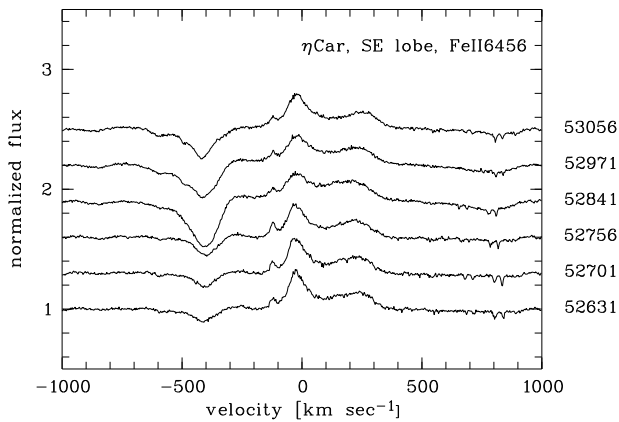}}
\caption{Same as Fig.~\ref{fig_P8_1D}, but for Fe\,{\sc
ii}$\lambda$6456.  The emission is relatively constant. The absorption
increases strongly at the event and remains strong for an extended
period. High-velocity absorption is not seen.}
\label{fig_feii_1D}
\end{figure}

Strong Fe\,{\sc ii} and [Fe\,{\sc ii}] lines are well-known features
in the direct spectrum of $\eta$~Car. In the reflected light, the
[Fe\,{\sc ii}] emission lines, narrow emission originating from
nebular structures close to $\eta$~Car, are much fainter than seen in
the direct spectrum of the star and close nebular structure.  We
discuss temporal variations seen by the FOS4 position
(Fig.~\ref{fig_feii_2D},~\ref{fig_feii_1D}) of the Fe\,{\sc
ii}$\lambda$6456 line here as representative of low ionization metal
lines.  The absorption in this line increases strongly (starting
around JD 2\,452\,800 - phase 0.990), but, in contrast to the hydrogen
lines, without a significant increase of the edge velocity.  The
strong absorption persists for an extended period of time, long after
the event, until the end of our monitoring. The profile of the
Fe\,{\sc ii} line is qualitatively similar to the Paschen lines, with
a pronounced red emission peak. The main peak is sharp and at a
velocity of about zero, the red peak is broader and peaks at about
$+220$ km sec${-1}$.

\begin{figure}[ht]
\resizebox{\hsize}{!}{\includegraphics[angle=270]{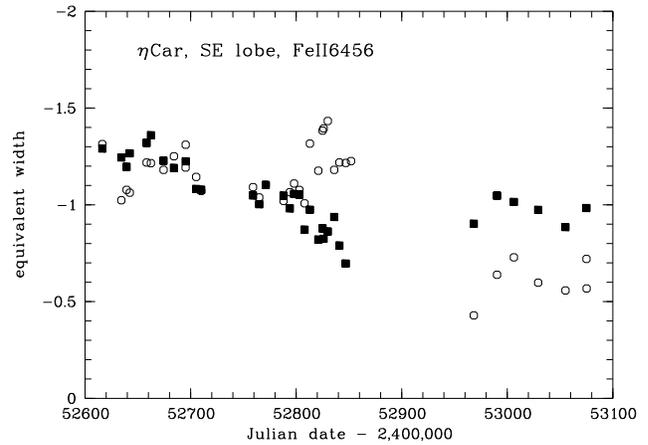}}
\caption{Equivalent width of Fe\,{\sc ii}$\lambda$6456 versus Julian
Date for position ``2'' (filled symbols) and position ``5'' (open
symbols). The equivalent width at ``5'' does not drop before the
event. However, after the event, the absorption is stronger than at position
``2''.}
\label{fig_feii_eqw}
\end{figure}

The time evolution of Fe\,{\sc ii}$\lambda$6456 changes strongly along
the slit position in the Homunculus. In Fig.~\ref{fig_feii_eqw} we
show the equivalent width variations at two different positions. While for
both positions the absorption strength is stronger after the event,
the effect is increases further out on the slit and thereby with angle
from the polar axis of the SE lobe. In addition, the
decrease begins at position ``2'' before the event (phase 0.97), 
but we do not see the decrease at position ``5''.
Only the fact that the equivalent width is less after the event
indicates a sharper and later drop in equivalent width.

\subsection{He\,{\sc i} lines}

\begin{figure}[ht]
\resizebox{\hsize}{!}{\includegraphics[angle=0]{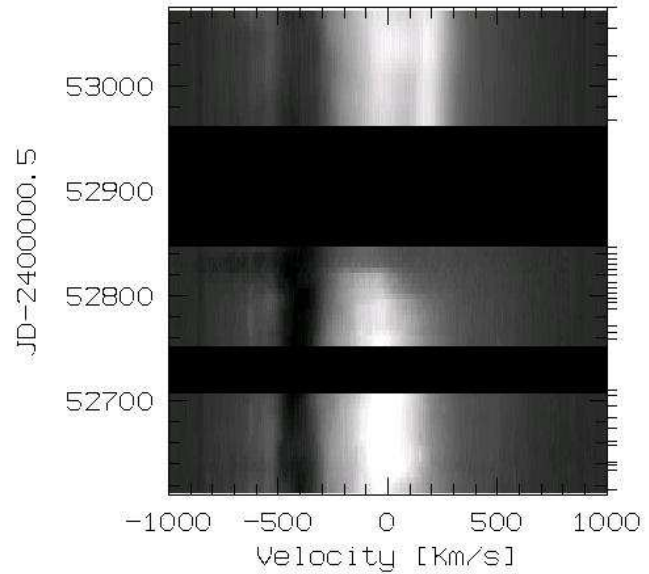}}
\caption{Same as Fig.~\ref{fig_P8_2D}, but for He\,{\sc i}$\lambda$6678.
The emission is strongly variable and nearly disappears at the
event. Note the brief episode of a high-velocity wind at the same
time. A red-shifted emission bump develops after the event.}
\label{fig_hei_2D}
\end{figure}

\begin{figure}[ht]
\resizebox{\hsize}{!}{\includegraphics[angle=270]{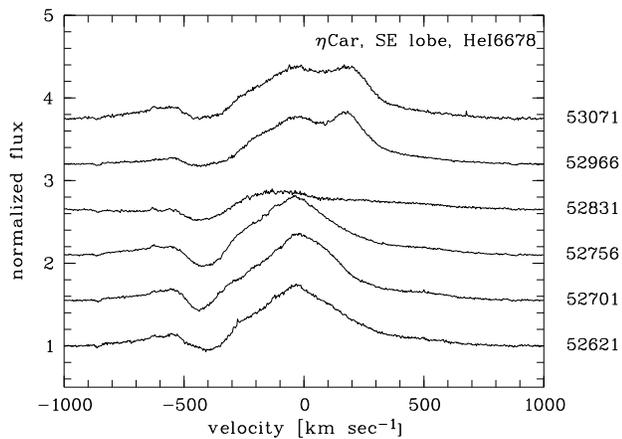}}
\caption{Same as Fig.~\ref{fig_P8_1D}, but for He\,{\sc
i}$\lambda$6678.}
\label{fig_hei_1D}
\end{figure}

The strong decrease in the strength of He\,{\sc i} emission is
characteristic for the event.  Fig.~\ref{fig_hei_2D},\ref{fig_hei_1D}
show time variations in the He\,{\sc i}$\lambda$6678 line.  It appears
that a flux minimum occurs around JD 2\,452\,844. In addition, a
brief episode of high-velocity absorption (edge velocity about $-750$
km sec$^{-1}$ at JD 2\,452\,830, corresponding to phase 0.005) is seen
in Fig.~\ref{fig_hei_2D}. The absorption strength decreases after the
event.

\begin{figure}[ht]
\resizebox{\hsize}{!}{\includegraphics[angle=0]{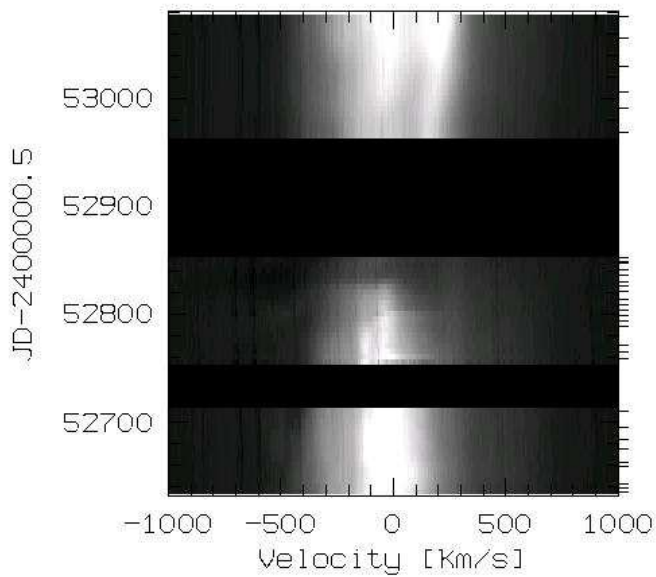}}
\caption{Same as Fig.~\ref{fig_P8_2D}, but for He\,{\sc
i}$\lambda$7065.  The emission is strongly variable and nearly
disappears at the event. In this line, absorption is always very weak,
except for a brief episode of a high-velocity wind at the event. A
red-shifted emission bump develops after the event.}
\label{fig_hei7065_2D}
\end{figure}

\begin{figure}[ht]
\resizebox{\hsize}{!}{\includegraphics[angle=270]{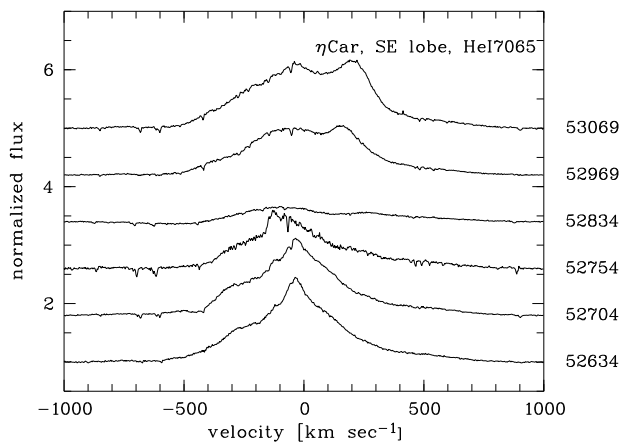}}
\caption{Same as Fig.~\ref{fig_P8_1D}, but for He\,{\sc
i}$\lambda$7065.}
\label{fig_hei7065_1D}
\end{figure}

Fig.~\ref{fig_hei7065_2D},\ref{fig_hei7065_1D} show the same for the
He\,{\sc i}$\lambda$7065 line. This line shows less absorption than
the He\,{\sc i}$\lambda$6678 line, but also this line shows evidence
for a brief episode of high-velocity absorption.

\begin{figure}[ht]
\resizebox{\hsize}{!}{\includegraphics[angle=270]{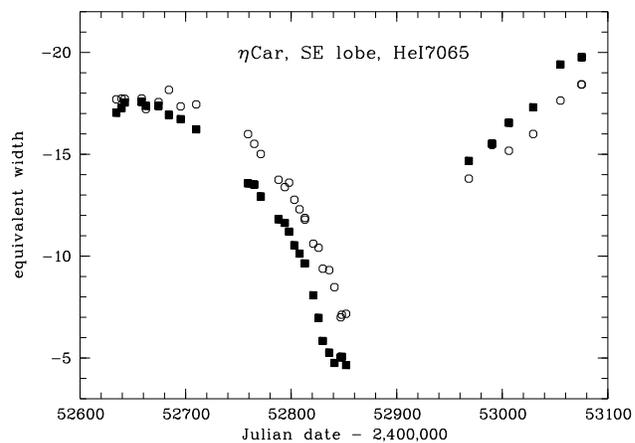}}
\caption{Equivalent width of He\,{\sc i}$\lambda$7065 versus Julian
Date for position ``2'' (filled symbols) and position ``5'' (open
symbols). The curve at position ``5'' is delayed by about 30 days and
the variations are somewhat smaller relative to position ``2''. }
\label{fig_hei_eqw}
\end{figure}

The evolution of the equivalent width of He\,{\sc i}$\lambda$7065 is
shown in Fig.~\ref{fig_hei_eqw}. The change in equivalent width is
dramatic. The equivalent widths outside of the event are similar at
position ``2'' and ``5'', but, as expected, the change is delayed
further out on the slit. However, the difference cannot be attributed
only to the time delay, since the curves are also intrinsically
different: The residual equivalent width appears smaller at position
``2'' (corresponding to light emitted close to the pole), which
indicates somewhat smaller changes there. This may be an artefact of
the observing gap, however, since the minimum at position ``5'' is
expected in this gap.

\subsection{He\,{\sc ii}$\lambda$4686}

\begin{figure}[ht]
\resizebox{\hsize}{!}{\includegraphics[angle=0]{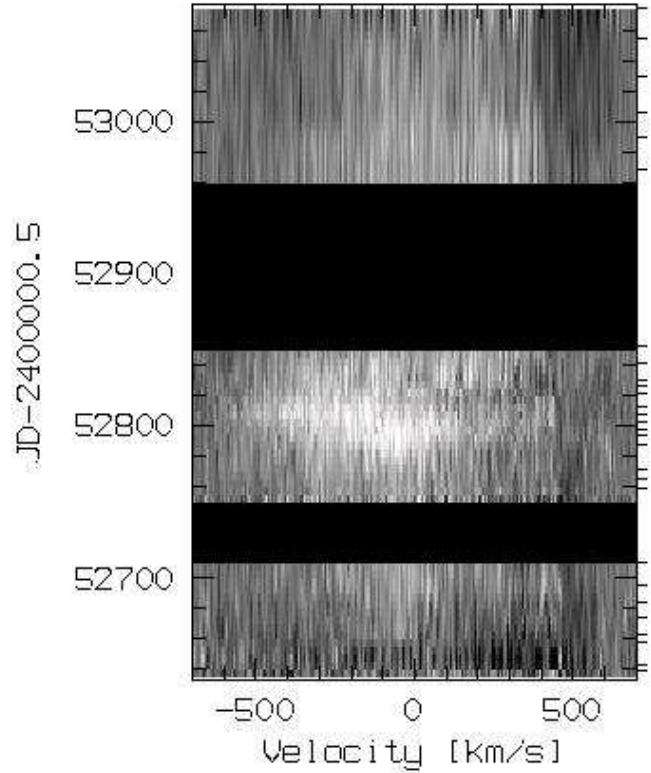}}
\caption{Same as Fig.~\ref{fig_P8_2D}, but for He\,{\sc
ii}$\lambda$4686.  Faint emission appears to be present well before
the event, but briefly increases strongly close to the event. The
emission disappears again just after phase zero. The emission is
blue-shifted and appears to move to the blue with time.}
\label{fig_heii_2D}
\end{figure}

\begin{figure}[ht]
\resizebox{\hsize}{!}{\includegraphics[angle=270]{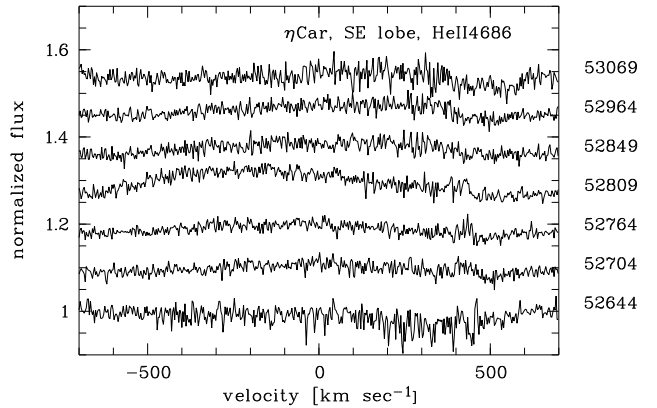}}
\caption{Same as Fig.~\ref{fig_P8_1D}, but for He\,{\sc
ii}$\lambda$4686.}
\label{fig_heii_1D}
\end{figure}

He\,{\sc ii}$\lambda$4686 is an important temperature diagnostic for
hot stars.  The absence of this line has been used by
\citet{2001ApJ...553..837H} to put constraints on the effective
temperature of $\eta$~Car. However, \citet{2004ApJ...612L.133S}
detected He\,{\sc ii}$\lambda$4686 in ground-based spectra of
$\eta$~Car for a brief episode close to the event.  Later, this line
was also detected by \cite{Martinetal2004} in STIS spectra centered on
the star. The He\,{\sc ii}$\lambda$4686 emission seems to originate in
a shock front. We find He\,{\sc ii}$\lambda$4686 also in the reflected
spectrum (cf.\,Figs.~\ref{fig_heii_2D},~\ref{fig_heii_1D}).

\begin{figure}[ht]
\resizebox{\hsize}{!}{\includegraphics[angle=270]{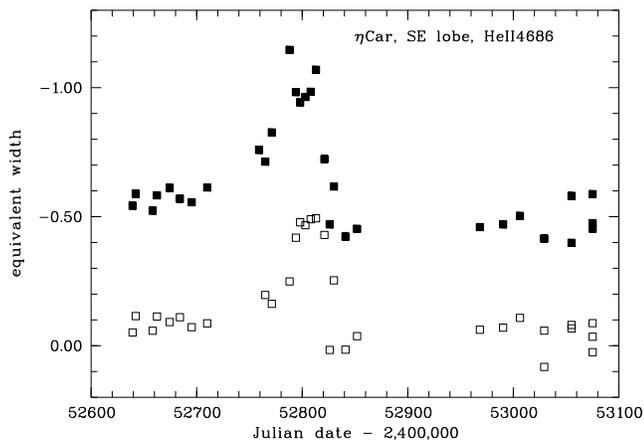}}
\caption{Equivalent width of He\,{\sc ii}$\lambda$4686 versus time as
measured at position ``2'' (open symbols) and closer to the center
(position ``0'', filled symbols).  The latter are shifted up to 0.5
units for clarity. The peak emission is stronger close to the center.}
\label{fig_heii_eqw}
\end{figure}

The emission is highly variable and clearly visible only for a very
brief period (about one month) around phase zero. It peaks around JD
2\,452\,815 (phase 0.998) and disappears again around JD 2\,452\,835
(phase 1.008). We do not detect this line clearly outside the
event. Surprisingly, the peak of the flux shows a time delay of only
about 10 days compared to the observations of
\citet{2004ApJ...612L.133S}, which have been taken centered on the
star.  The observations of \cite{Martinetal2004} show the highest
equivalent width almost exactly at the same date as our spectra. In
their spectra, the equivalent width is, however, almost zero at
2\,452\,825, while we still measure a significant equivalent width at
2\,452\,830. Because of the special importance of this line, we also
measured its equivalent width close to the star (see below). The
measured values at position ``2'' and ``0'' are shown in
Fig.~\ref{fig_heii_eqw}. This plot shows that the rise and fall of the
emission is reasonably well sampled with our spectra. The emission
feature is broad and faint, therefore the S/N in the equivalent width
measurement is quite low. At larger distance the S/N is still lower
and the emission can barely be measured. We conclude that the peak of
our measurements is somewhat delayed compared to direct spectra, but the
delay appears to be limited to about ten days.

As observed in spectra of the center, the He\,{\sc ii} emission is
significantly blue-shifted. The maximum equivalent width observed is
about 500 m\AA, about a factor of two lower than in ground-based
spectra of the central source \citep{2004ApJ...612L.133S} and still
smaller than in STIS spectra taken directly on the star
\citep{Martinetal2004}.

At positions further out on the slit, the emission is also present,
but appears, as expected, later in the cycle. It should be stressed,
however, that the sampling of our spectra is insufficient to determine
the exact date of the peak in He\,{\sc ii} emission better than within
about two weeks.

\begin{figure}[ht]
\resizebox{\hsize}{!}{\includegraphics[angle=0]{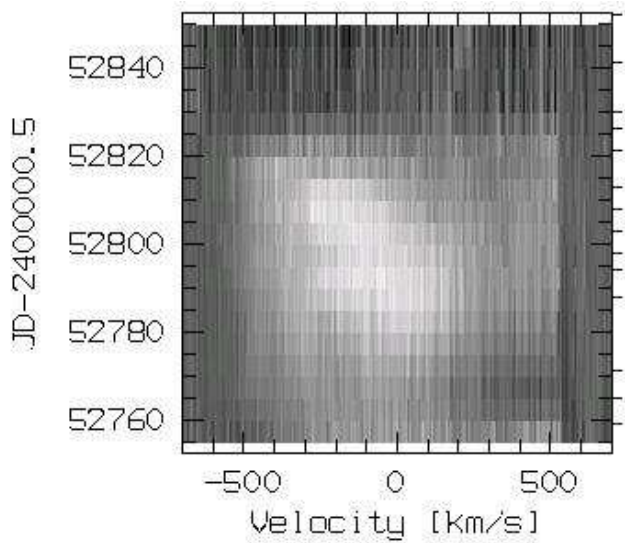}}
\caption{Same as Fig.~\ref{fig_P8_2D}, but for He\,{\sc
ii}$\lambda$4686 extracted close to the center of the Homunculus. The
emission is blue-shifted and appears to move to the blue with time. The
apparent absorption at the right side is a artefact due to the order
merging. This is a close-up covering only the well-sampled part close
to the event. The tick marks at the right side show the dates where
spectra have been taken.}
\label{fig_heii_2D_0}
\end{figure}

Since He\,{\sc ii}$\lambda$4686 has a high diagnostic value, we also
extracted the spectrum closer to the central source. The dynamical
spectrum extracted 2\farcs5 inwards from the center of the slit is
shown in Fig.~\ref{fig_heii_2D_0}. Here the blue-ward motion of the
emission peak is clearer than in Fig.~\ref{fig_heii_2D}, possibly as
a result of higher S/N. This shift has already been found and
discussed by \citet{2004ApJ...612L.133S}. The tick marks at the right
side in Fig.~\ref{fig_heii_2D_0} show the dates where spectra have
been taken. This shows that the peak is reasonably well sampled, but
obviously some data are interpolated. The equivalent widths are shown
in Fig.~\ref{fig_heii_eqw}. The peak is stronger closer to the center
and also peaks earlier by about 10 days.

\section{Discussion and conclusion}

Our UVES observations show in significant detail the evolution of line
profiles during a spectroscopic event in reflected light. 
Assuming that the emission observed at FOS4 is only a scattered
spectrum originating from the central source and does not include
emission from other sources, these observations correspond to an
about pole-on view of $\eta$~Car, while the direct observations of the
star see it at about 45$^\circ$. While many changes are qualitatively
similar to what was already known from observations of the central
source, there are some important differences. We have also
significantly better temporal sampling, spectral resolution and S/N
than \citep{2003ApJ...586..432S}.

The Balmer lines have already been discussed in detail by
\citet{Davidsonetal2005} and \citet{Weisetal2005}. The decrease of the
equivalent width of the hydrogen lines during the spectroscopic event
is pronounced, but not as strong as seen in the direct line of sight
\citep{Davidsonetal2005}. The equivalent width is always significantly
smaller than observed directly on the star.

The appearance of a strong red bump in the reflected spectra has
already been described for the Balmer lines by \citet{Weisetal2005}.
We see the same feature in many different lines, in particular after
the event. This feature is not a peculiarity of the present
observations, but also present in spectra obtained in Dec.~1999 during
UVES commissioning \citep{Weisetal2005}. At present, the origin of
this component and its relation with the spectroscopic cycle is not
clear.

The changes in the high-excitation lines such as He\,{\sc i} are
particularly pronounced. The equivalent widths change dramatically
and, in addition, a high-velocity absorption appears very briefly near
the event.  High-velocity absorption has also been observed in
H$\alpha$ \citep{Weisetal2005}, but since this absorption appears long
after the spectroscopic event, it is not clear if it is related to the
event.

The low-ionization metal lines such as Fe\,{\sc ii} show, similar to the
hydrogen lines, an increasing absorption during the event, but no evidence for
high-velocity absorption. For these lines, the recovery phase after the event
seems particularly extended. These lines also seem to have a complicated 
dependence on the viewing angle.

He\,{\sc ii}$\lambda$4686 line emission is seen in the spectrum of the SE
lobe, consistent with viewing the polar region, but with significantly smaller
equivalent width than seen of the star directly. If corrected for light-travel
time delay, the polar emission of the He\,{\sc ii}$\lambda$4686 appears to
peak earlier than closer to the equator (45$^\circ$), constraining models for
the origin of this emission. 

The He\,{\sc ii}$\lambda$4686 line is only one example of how the geometry of the
region where the event occurs can be constrained by comparing the timing of
the event as seen in reflected light in the Homunculus and in direct spectra.
In practice this is complicated by several factors:

\begin{itemize} 
\item The observations of the reflected spectra have to be
  corrected for the light-travel time delay. 
\item With the UVES data, only the beginning of the event is well sampled. The
  recovery phase is sampled incompletely because of observing constraints.
\item The STIS data are also not very densely sampled. 
\item The qualitative temporal behaviour of e.g.\ the equivalent widths is
  significantly different between positions in the SE lobe of the Homunculus
  and direct spectra. This makes it difficult to derive a simple timing
  difference.
\end{itemize} 

Only detailed quantitative modeling of the event can at least partly resolve
these problems. However, already the equivalent width variations measured in
UVES spectra of the Homunculus at different slit positions clearly indicate
that the event depends on the viewing angle. 

The spectral changes of the hydrogen lines and even more the He\,{\sc
i} and Fe\,{\sc ii} lines are clearly not symmetric in time. The
``ingress'' phase is more rapid than the ``egress'' phase. The changes
can be described as a ``shell-like'' event, i.e.\ a fast onset and a
slow recovery phase.  Such a shell-like event is especially supported
by a new feature not described before: A phase of high-velocity
absorption, which appears very briefly near the event especially in
the He\,{\sc i} lines. This is one of the strongest indications that
the spectroscopic event involves some mass ejection and is not simply
due an eclipse event. 

A more detailed comparison of the reflected spectra taken at different slit
positions and with the STIS spectra obtained directly at the center will be
the subject of following papers.

\begin{acknowledgements}
It is a pleasure to thank the ESO Paranal Observatory staff for
carrying out for us the service mode observations on which this paper
is based.
\end{acknowledgements}

\bibliographystyle{aa}
\bibliography{etacar}

\end{document}